\documentclass[a4paper,fleqn]{cas-dc}

\usepackage[numbers]{natbib}

\def\tsc#1{\csdef{#1}{\textsc{\lowercase{#1}}\xspace}}
\tsc{WGM}
\tsc{QE}

\begin{document}
\let\WriteBookmarks\relax
\def\floatpagepagefraction{1}
\def\textpagefraction{.001}

\shorttitle{Local lattice dynamics of hcp zinc from EXAFS and machine-learning  interatomic potentials}    

\shortauthors{V. Dimitrijevs et al.}  

\title [mode = title]{Local lattice dynamics of hcp zinc from EXAFS and machine-learning interatomic potentials}  



%

\author[1]{Vitalijs Dimitrijevs}[orcid=0000-0002-3892-9857]


\ead{vitalijs.dimitrijevs@cfi.lu.lv}


\credit{Investigation,  writing -- original draft, writing -- review \& editing}

\affiliation[1]{organization={Institute of Solid State Physics, University of Latvia},
	addressline={Kengaraga street 8}, 
	city={Riga},
	postcode={LV-1063}, 
	country={Latvia}}

\author[1]{Pjotrs \v{Z}guns}[orcid=0000-0002-3387-2919]


\ead{pjotrs.zguns@cfi.lu.lv}


\credit{Investigation, software, writing—original draft, writing – review \& editing}

\author[1]{Inga Pudza}[orcid=0000-0002-8692-3089]


\ead{inga.pudza@cfi.lu.lv}


\credit{Investigation, writing—original draft, writing – review \& editing}


\author[1,2]{Aleksandr Kalinko}[orcid=0000-0002-0615-4735]


\ead{aleksandr.kalinko@desy.de}


\credit{Funding acquisition, investigation, project administration, writing—original draft, writing – review \& editing }

\affiliation[2]{organization={Deutsches Elektronen-Synchrotron DESY},
	addressline={Notkestr. 85}, 
	city={Hamburg},
	postcode={22607}, 
	country={Germany}}

\author[1]{Alexei Kuzmin}[orcid=0000-0003-4641-6354]

\cormark[1]


\ead{a.kuzmin@cfi.lu.lv}


\credit{Conceptualization, methodology, investigation, writing—original draft, writing – review \& editing, supervision}

\cortext[cor1]{Corresponding author}



\begin{abstract}
The lattice dynamics of hexagonal close-packed (hcp) zinc, a prototypical anisotropic metal, is studied using temperature-dependent Zn K-edge extended X-ray absorption fine structure (EXAFS) spectroscopy combined with atomistic simulations. The reverse Monte Carlo method enable the extraction of mean-square relative displacements (MSRDs) for eight coordination shells, providing a shell-resolved description of thermal motion. The MSRD temperature dependence, analysed using the correlated Einstein model, yields effective interatomic force constants and reveals pronounced anisotropy between in-plane and out-of-plane interactions. This anisotropy is further quantified by the ratio of MSRDs for the first and second coordination shells, which closely matches the anisotropic displacement parameters from diffraction experiments. Molecular dynamics simulations using the CHGNet universal machine-learning interatomic potential show that the original model overestimates thermal disorder, while a fine-tuned version substantially improves agreement with experimental EXAFS spectrum and radial distribution function. Overall, EXAFS-informed analysis is effective for validating and refining machine-learning interatomic potentials.
\end{abstract}

\begin{graphicalabstract}
\includegraphics[width=.5\columnwidth]{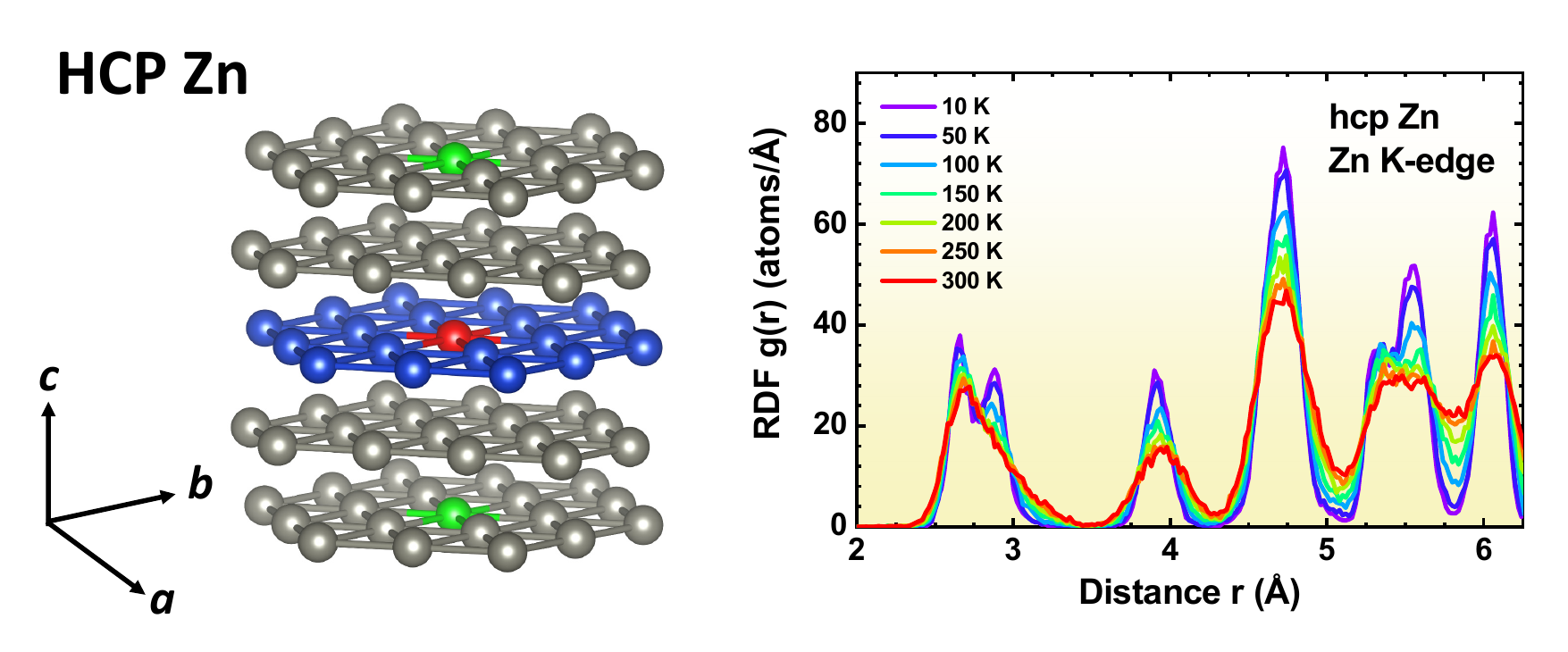}
\end{graphicalabstract}

\begin{highlights}
\item Lattice dynamics of hcp zinc was studied using EXAFS spectroscopy.
\item Mean-square relative displacements for the eight coordination shells were obtained.
\item Pronounced anisotropy between in-plane and out-of-plane interactions was revealed. 
\item Molecular dynamics simulations were performed using machine-learning interatomic potential.
\item Fine-tuning the potential improves agreement with the experiment.
\end{highlights}



\maketitle

\section{Introduction}

The hexagonal close-packed (hcp) structure is one of the most efficient ways of packing atoms in a solid and is characterized by the  $c/a$\  ratio of the lattice parameters of the unit cell.
The space group of the hcp lattice is $P6_3/mmc$ (No. 194) with two atoms per
unit cell located in the special Wyckoff positions 2$c$	 (1/3, 2/3, 1/4) \cite{Jette1935}.  
In the ideal hcp structure, the $c/a$\  ratio is equal to 1.633 and deviates from this value by no more than 4\% for almost all elements (e.g., Mg, Ti, Co, Zr). However, there are two exceptions, zinc and cadmium, for which the $c/a$\  ratio value (1.86 for zinc and 1.89 for cadmium) is significantly larger than the ideal value by about 15\% \cite{Wyckoff1963}.
The larger $c/a$ ratio stretches the lattice along the $c$-axis making it highly anisotropic that ultimately affects the material properties. 

\begin{figure}
	\centering
	\includegraphics[width=.9\columnwidth]{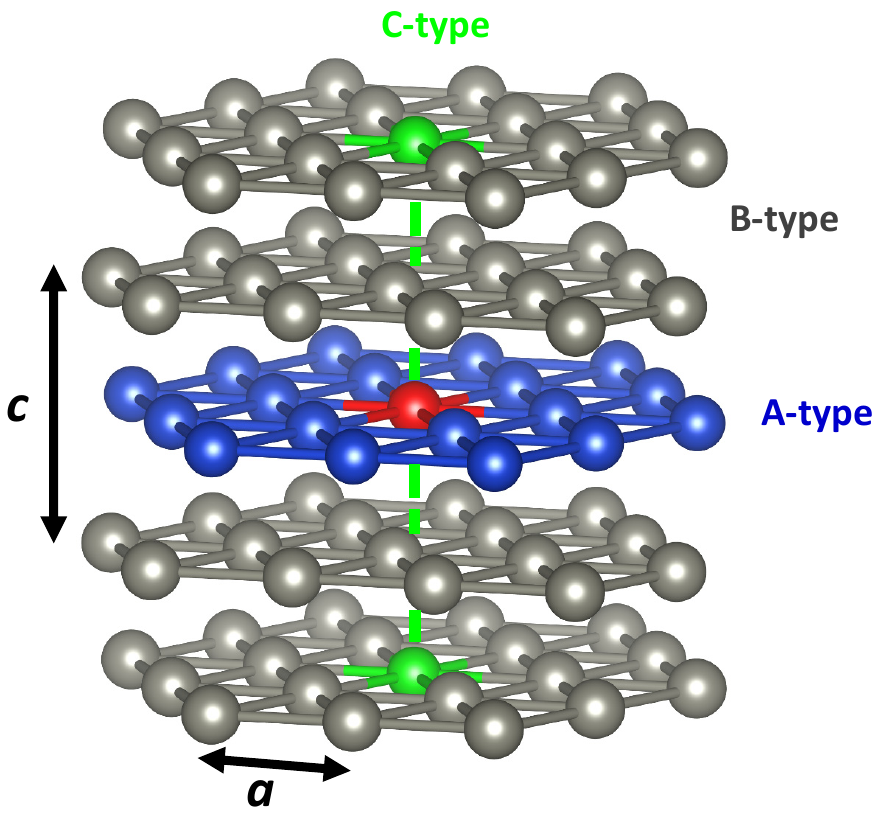}	
	\caption{Fragment of the crystal structure of hcp zinc showing the stacking sequence of layers and the lattice parameters $a$ and $c$. Chemical bonds between nearest-neighbor zinc atoms are shown. The absorbing zinc atom is highlighted in red. Two zinc atoms located above and below the absorbing zinc (C-type) are highlighted in green. Zinc atoms in the same plane as the absorbing zinc (A-type) are highlighted in blue. Other zinc atoms (B-type) are shown in gray. The vertical dashed line serves as a guide to the eye connecting the two C-type zinc atoms.}
	\label{fig1}
\end{figure}

The strong anisotropy of zinc structure  (Fig.\ \ref{fig1}) results in an anisotropic potential field where interatomic interactions and thermal vibrations differ drastically between the basal plane and the hexagonal $c$-axis \cite{Nuss2010}. 
Because of the high $c/a$\ ratio, zinc atoms are more closely packed within the hexagonal layers ($ab$ plane) than between them.  This creates an asymmetric potential environment that cannot be accurately described by harmonic (quadratic) terms alone thus leading to anharmonic effects \cite{Barron1967,Skelton1968,Albanese1976,Tomaschitz2021}.

The studies of the lattice dynamics in hcp zinc attracted continuous attention for years \cite{Nuss2010,Barron1967,Skelton1968,Albanese1976,Merisalo1977,Rossmanith1978,Merisalo1979,Pathak1981,Peng1996,Hung2008,Masadeh2019,Masadeh2022}. 
Accurate single-crystal X-ray diffraction studies \cite{Nuss2010} performed in the temperature range of 40 to 500 K showed that the thermal expansion of zinc along $c$ axis is about 10 times that along $a$ axis and the thermal motion of Zn atoms is strongly anisotropic. Specifically, the ratio of anisotropic displacement parameters \cite{Trueblood1996} for out-of-plane vibrations ($U_{33}$) and in-plane vibrations ($U_{11}$), i.e. the ratio $U_{33}/U_{11}$, is highly sensitive to temperature, increasing from  $\approx$1.6 at 40~K to $\approx$2.6 at 500~K.  These findings were also supported by more recent  X-ray total scattering experiments \cite{Masadeh2019,Masadeh2022}. 

X-ray absorption spectroscopy (XAS) is a local, element-specific tool well suited for studying dynamic structural disorder and anharmonic lattice physics in materials. The analysis of temperature-dependent extended X-ray absorption fine structure (EXAFS) spectra using the cumulant expansion provides a natural framework for quantifying such anharmonic effects \cite{Bunker1983,Fornasini2005}. Moreover, EXAFS-based analyses using model potentials have demonstrated their power in providing dynamical information about local lattice instabilities that are inaccessible to conventional crystallographic measurements \cite{MustredeLeon1992}. However, it should be noted that these approaches are designed for and validated on the first one or two coordination shells only. Their applicability to distant shells 	is severely limited by the proliferation of multiple-scattering paths, increased thermal disorder causing poor convergence of the cumulant series, and the difficulty of isolating individual shells by Fourier filtering.

In this study, temperature-dependent (10-300~K) XAS, combined with reverse Monte Carlo (RMC) analysis, is employed to probe the anisotropy of local lattice vibrations in hcp zinc. The RMC approach adopted here utilizes a large supercell with periodic boundary conditions, in which multiple scattering contributions are explicitly calculated for each atomic configuration, while all coordination shells are treated simultaneously. This approach enables the reliable and self-consistent extraction of mean-square relative displacements (MSRDs) and effective force constants for up to eight coordination shells. As a result, we are able to quantify the anisotropy of thermal vibrations and evaluate the ability of modern interatomic potentials to reproduce these effects. To this end, molecular dynamics (MD) simulations based on universal machine-learning interatomic potentials (uMLIPs) are employed.

\section{Experimental and calculations}

\subsection{X-ray absorption experiments}

Temperature-dependent XAS experiments were conducted at the DESY PETRA-III storage ring, 
which operated at $E$ = 6.08~GeV and $I$ = 120~mA in the top-up mode with 480 bunches. 
The  Zn K-edge (9659~eV) X-ray absorption spectra of hcp zinc foil (99.9\%, Goodfellow Cambridge Ltd. ) with a thickness of 5~$\mu$m  were collected at the P65 Applied X-ray Absorption Spectroscopy undulator beamline \cite{Welter2019} in transmission mode, using a fixed-exit Si(111) monochromator and two ionization chambers filled with nitrogen gas. Two uncoated Si plane mirrors were employed for harmonic reduction. 
Measurements were performed in continuous scan mode at seven temperatures (10, 50, 100, 150, 200, 250, and 300~K) using a Janis STVP-FTIR  He-flow cryostat (Janis Research Company, LLC, Woburn, MA, USA) and a Lake Shore Model 335 temperature controller  (Lake Shore Cryotronics, Inc., Woburn, MA, USA). 

\begin{figure}
	\centering
	\includegraphics[width=.9\columnwidth]{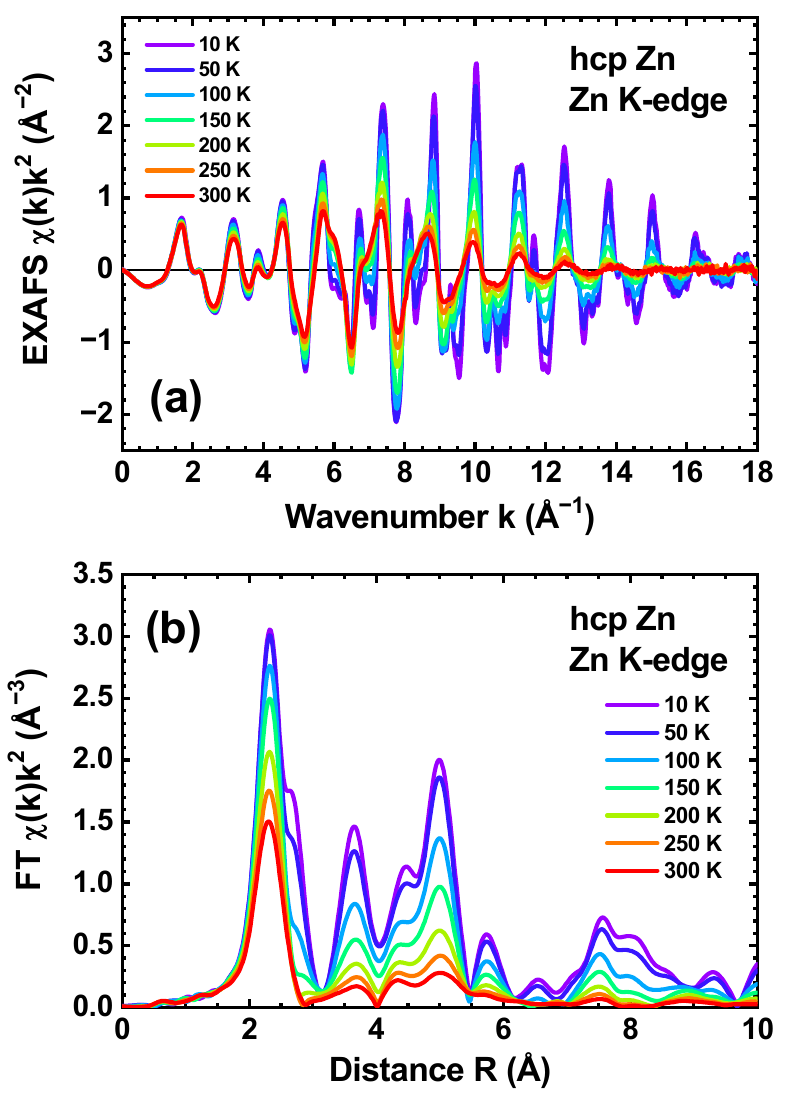}	
	\caption{Experimental Zn K-edge EXAFS spectra  $\chi(k)k^2$ (a) and their Fourier transforms (FTs) (b) for hcp zinc as a function of temperature.  Only the moduli of the FTs are shown. }
	\label{fig2}
\end{figure}

The Zn K-edge EXAFS spectra $\chi(k)k^2$\ (where $k$ is the photoelectron wavenumber, defined as $k = \sqrt{(2m_e/\hbar^2)(E-E_0)}$, with $m_e$ being the electron mass,  $\hbar$ the reduced Planck constant, and $E_0$ the threshold energy)  were extracted over a wide $k$-space range up to $k_{\rm max}$ = 18~\AA$^{-1}$\ using the XAESA code \cite{XAESA}, following the conventional procedure \cite{Kuzmin2014}. The experimental EXAFS spectra  and their Fourier transforms (FTs) as a function of temperature are shown in Fig.\ \ref{fig2}. The FTs were calculated using a 10\% Gaussian function. Note that all peak positions in the FTs are shifted to smaller distances relative to their crystallographic values because the FTs were not corrected for the atomic backscattering phase shift. 

\subsection{Reverse Monte Carlo simulations} 
\label{RMC}

The experimental Zn K-edge EXAFS spectra $\chi(k)k^2$\ were analysed using the RMC method enhanced with an evolutionary algorithm (EA), as implemented in the EvAX code and described in detail in \cite{Timoshenko2012rmc,Timoshenko2014rmc}. The structural model (simulation box) of hcp zinc was  a  6$a$ $\times$ 6$a$ $\times$ 4$c$ supercell with periodic boundary conditions. This box included 288 zinc atoms and  was constructed considering the temperature dependence of the lattice parameters $a$ and $c$, as obtained from an accurate single-crystal X-ray diffraction experiment in \cite{Nuss2010}. Thirty-two atomic configurations were simultaneously used in the EA algorithm to expedite the optimization procedure \cite{Timoshenko2014rmc}.  The atoms were allowed to displace from their equilibrium positions in the ideal hcp zinc structure by up to 0.4~\AA\ to account for thermal disorder. 

At each step of the RMC/EA simulation, configuration-averaged (CA) EXAFS spectra were evaluated over all absorbing atoms in the simulation box.  The ab initio real-space multiple-scattering FEFF8.5L code \cite{Rehr2000,Ankudinov1998} was used to calculate the EXAFS spectrum for each atom, employing the complex energy-dependent exchange--correlation Hedin--Lundqvist potential \cite{Hedin1971} to account for inelastic effects. Multiple-scattering effects up to the fourth order were included in the calculations \cite{Zabinsky1995,Ravel2005,Rehr2009}.

\begin{figure*}
	\centering
	\includegraphics[width=.9\textwidth]{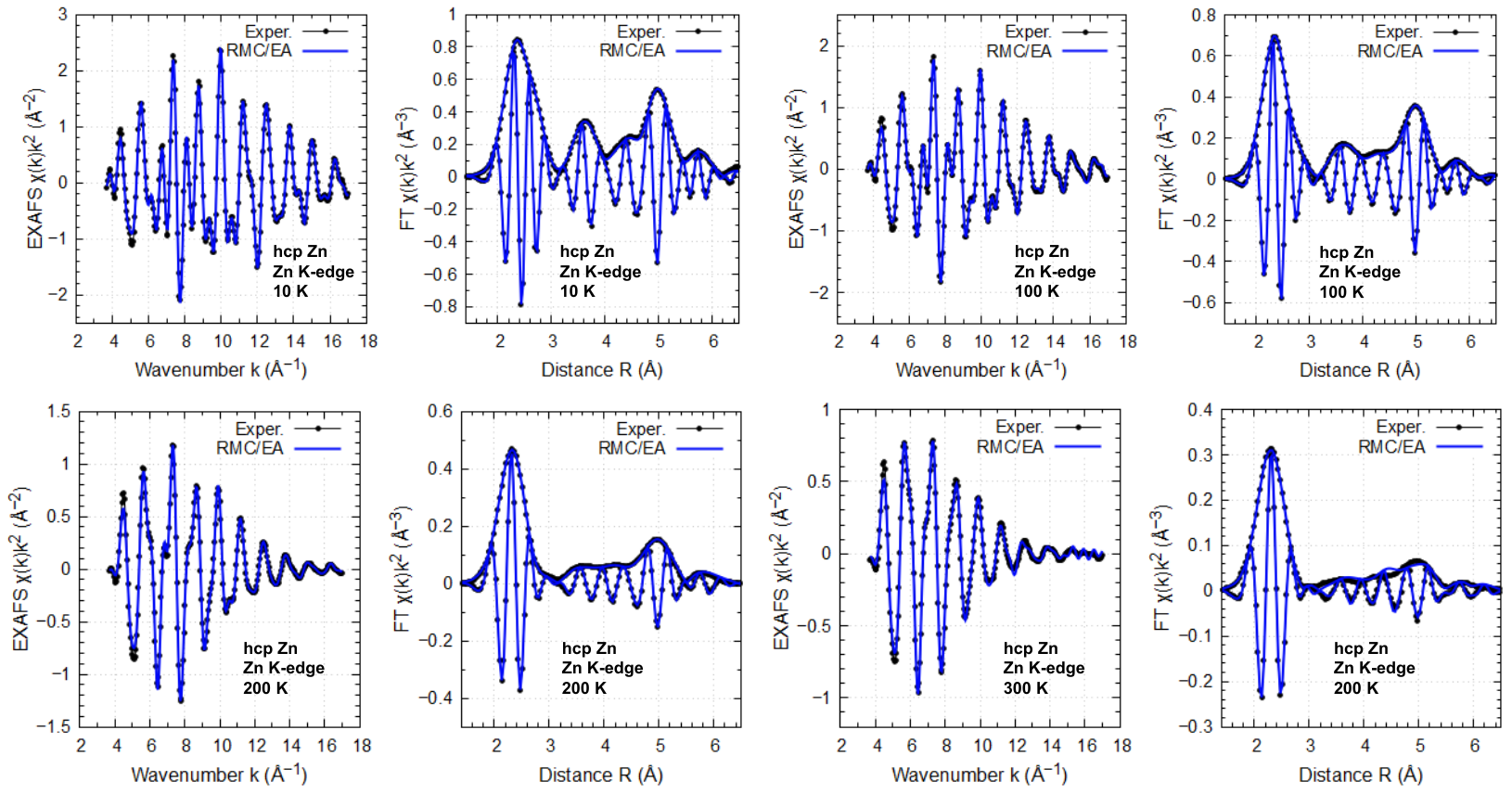}	
	\caption{Experimental (solid circles) and RMC/EA-simulated (solid lines) Zn K-edge EXAFS spectra $\chi(k)k^2$ and their Fourier transforms (FTs) for hcp zinc at four selected temperatures (10, 100, 200, and 300~K). Both the moduli and imaginary parts of the FTs are shown. }
	\label{fig3}
\end{figure*}

The difference between the Morlet wavelet transforms (WTs) \cite{Timoshenko2009wt} of the experimental and calculated CA-EXAFS spectra was minimized simultaneously in $k$-space from 3.7~\AA$^{-1}$\ to 17.0~\AA$^{-1}$\  and in $R$-space from 1.4~\AA\ to 6.5~\AA\  \cite{Timoshenko2014rmc}. Good agreement was achieved  at all temperatures, with results at four selected temperatures (10, 100, 200, and 300~K) shown in  Fig.\ \ref{fig3}.
RMC/EA simulations were repeated ten times at each temperature using different sequences of pseudo-random numbers for atomic displacements to obtain reasonable statistics. 
The atomic coordinates in the final configurations were used to calculate the mean radial distribution functions (RDFs) $g_\text{Zn--Zn}(r)$, as shown in Fig.\ \ref{fig4}.  These RDFs were decomposed into a set of Gaussian peaks (see Supplementary Material) to extract the MSRDs $\sigma^2$\ for Zn--Zn atom pairs within the nearest nine coordination shells.

The RMC/EA analysis was also repeated using $k^3$ weighting for two extreme temperatures (10 and 300~K). The resulting EXAFS fits, RDFs, and MSRDs are compared with those obtained using the $k^2$ weighting in Figs.\ S2–S5 (see Supplementary Material). The structural parameters derived from the two analyses are in very good agreement. In particular, the shapes of the RDFs and the MSRD values for the nine coordination shells of zinc remain essentially unchanged within the uncertainties of the method, demonstrating the robustness of the results with respect to the choice of EXAFS weighting and and showing that they are not biased by the reduced contribution of the high-$k$ region at elevated temperatures.

\begin{figure}
	\centering
	\includegraphics[width=.9\columnwidth]{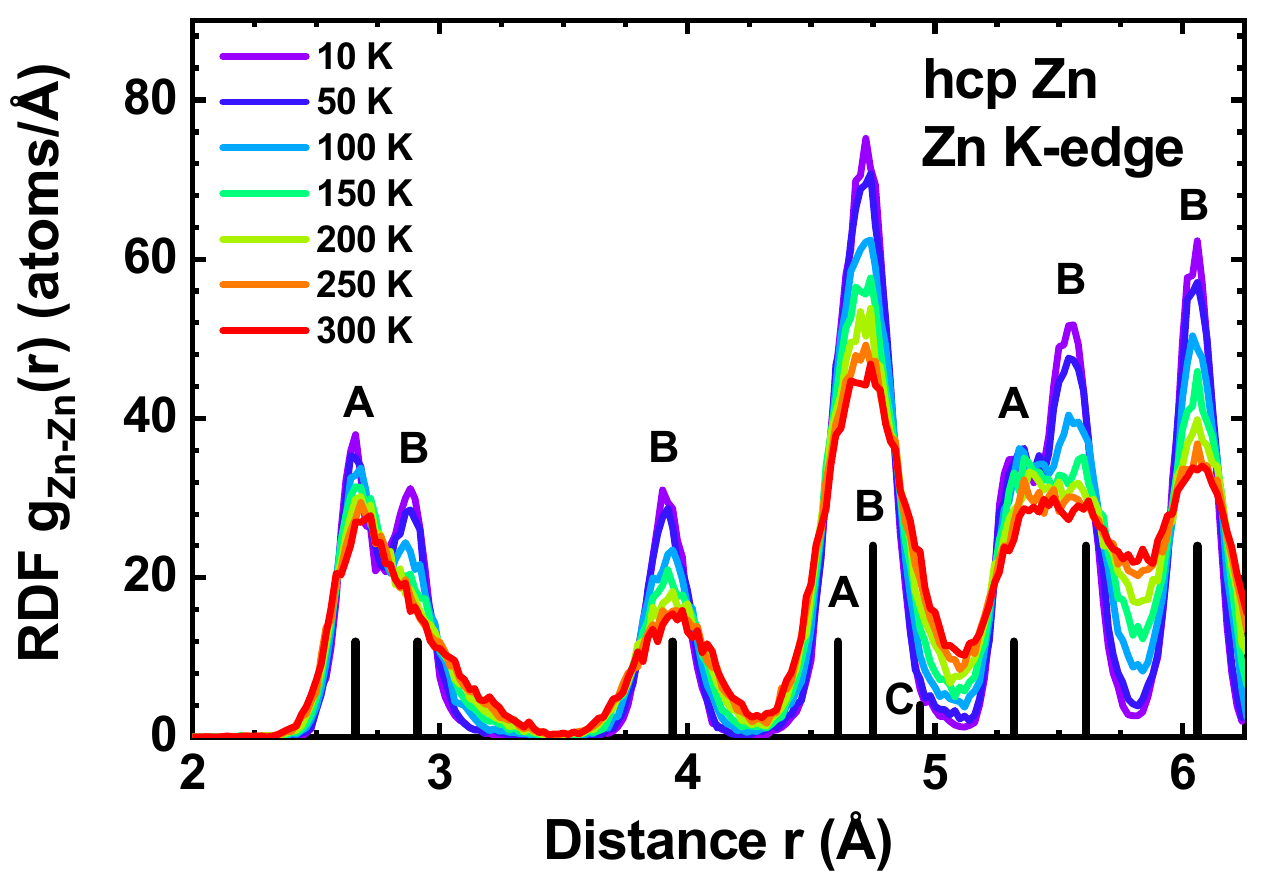}
	\caption{Temperature dependence of the radial distribution functions (RDFs) $g_\text{Zn--Zn}(r)$\ in hcp zinc obtained from RMC/EA simulations. Vertical bars indicate the positions of the coordination shells as determined by diffraction. The shells are labeled by atom type according to Table\ \protect\ref{table1}.}
	\label{fig4}
\end{figure}

\begin{figure}
	\centering
	\includegraphics[width=.9\columnwidth]{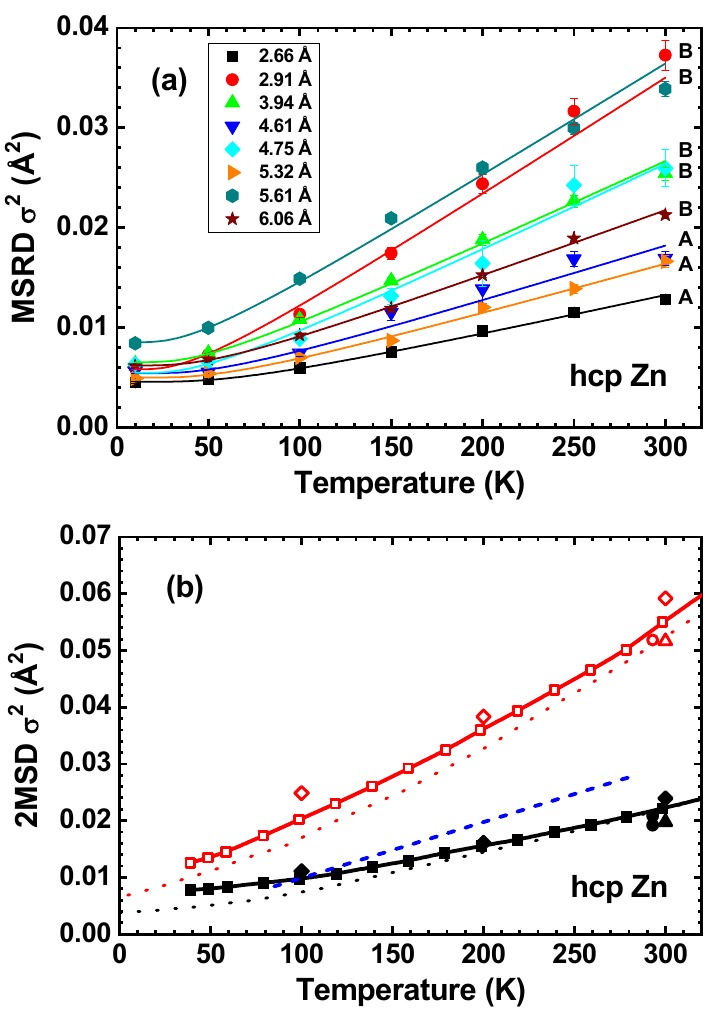}	
	\caption{(a) Temperature dependence of the mean-square relative displacements (MSRDs) for Zn--Zn atom pairs in eight coordination shells, obtained from the Zn K-edge EXAFS spectra in this study and labeled by atom type.	 
		Solid lines in (a) represent fits using the Einstein model with the effective force constants ($\kappa$) reported in Table\ \protect\ref{table1}. 
		(b) Temperature dependence of the mean-square displacements (MSDs), obtained by single-crystal X-ray diffraction in \protect\cite{Skelton1968} (dotted lines),  \protect\cite{Rossmanith1978} (circles) and \protect\cite{Nuss2010} (squares), from PDF (diamonds) and Rietveld (triangles) refinements reported in \protect\cite{Masadeh2022}. Solid and open symbols correspond to in-plane ($2 U_{11}$) and out-off-plane ($2 U_{33}$) MSDs, respectively. The dashed line shows 2MSD($T$) calculated in \protect\cite{Peng1996} from the experimentally determined phonon density of states. }
	\label{fig5}
\end{figure}

\begin{figure}
	\centering
	\includegraphics[width=.9\columnwidth]{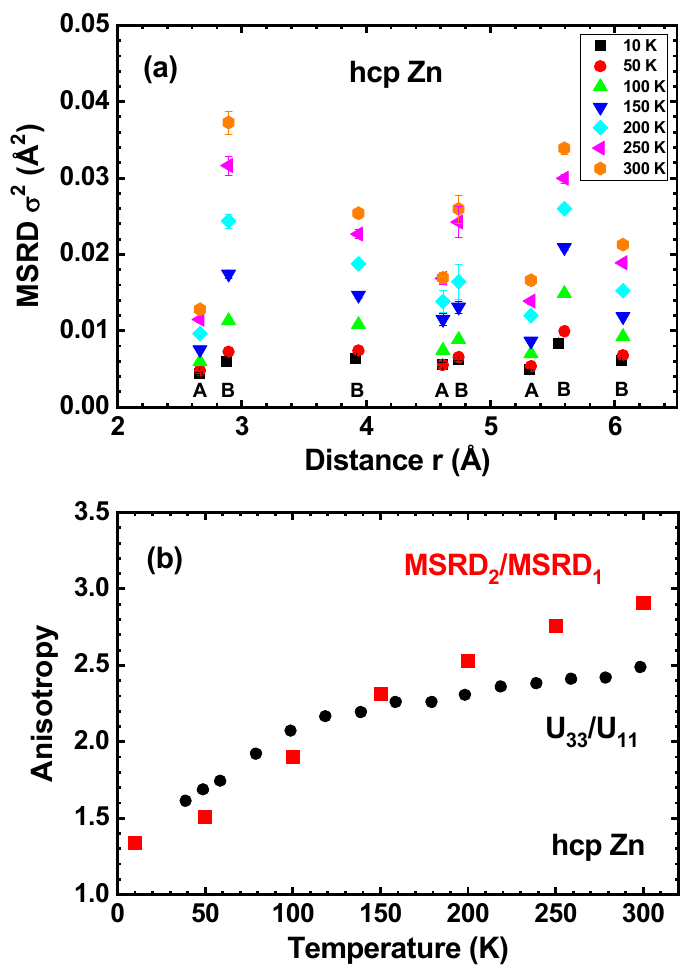}	
	\caption{(a) Dependence of the mean-square relative displacements (MSRDs) for Zn--Zn atom pairs on interatomic distance at seven temperatures.  The coordination shells are labeled by atom type according to Table\ \protect\ref{table1}.
		(b) Temperature dependence of the anisotropy, expressed by the ratio of  MSRD$_{2}$/MSRD$_{1}$  for the  second (MSRD$_2$) and first (MSRD$_1$) coordination shells (squares), and by the ratio of the  out-off-plane ($U_{33}$) and in-plane ($U_{11}$)  MSD parameters  $U_{33}/U_{11}$ (circles) from \protect\cite{Nuss2010}.} 
	\label{fig6}
\end{figure}

The temperature and interatomic distance dependences of the MSRDs are shown in Figs.\ \ref{fig5}(a) and \ref{fig6}(a)
for eight coordination shells, corresponding to A-type and B-type zinc atoms (Fig.\ \ref{fig1}).  
Note that the results for two C-type zinc atoms located above and below the absorbing zinc had high uncertainty 
due to poor statistics and, therefore, are not reported. 
The correlated Einstein model \cite{Sevillano1979} was fitted to the temperature dependences of the MSRDs (Figs.\ \ref{fig5}(a)) to determine the effective force constants ($\kappa$) (Table\ \ref{table1}). 

\begin{table}[]
	\caption{Atom types (see Fig.\ \protect\ref{fig1}),   crystallographic coordination numbers ($N$),  and Zn--Zn interatomic distances ($r$) for the first nine coordination shells of Zn in hcp zinc. 
		The effective force constants ($\kappa$) were obtained using the correlated Einstein model \protect\cite{Sevillano1979} from the temperature dependences of the MSRDs for each shell. }   
	\label{table1}
	\begin{tabular*}{\tblwidth}{@{} LLLLL@{} } 
		\toprule
		Shell   & Atom type& $N$  & $r$ (\AA) & $\kappa$ (N/m)  \\
		\midrule
		1   & A & 6   & 2.66 &  34 $\pm$ 1  \\
		2   & B & 6   & 2.91 &  12 $\pm$ 1  \\		
		3   & B & 6   & 3.94 &  16 $\pm$ 1  \\		  		
		4   & A & 6   & 4.61 &  24 $\pm$ 2  \\	  
		5   & B & 12  & 4.75 &  16 $\pm$ 1   \\					
		6   & C & 2   & 4.94 &  -   \\
		7   & A & 6   & 5.32 &  27 $\pm$ 1  \\		
		8   & B & 12  & 5.61 &  12 $\pm$ 1  \\		
		9   & B & 12  & 6.06 &  21 $\pm$ 1  \\		
		\bottomrule
	\end{tabular*}
\end{table}

\subsection{Molecular Dynamics Simulations}
\label{MD}

Molecular Dynamics (MD) simulations were performed in the canonical (NVT) ensemble
using the Nos\'{e}--Hoover thermostat (with a time constant of 30 fs) to maintain a temperature of 300~K. A timestep of 1~fs was applied to integrate the equations of motion,
as implemented in the Atomistic Simulation Environment \cite{ase_paper}.
The  CHGNet universal machine learning interatomic potential (uMLIP) (version 0.3.0) \cite{deng2023chgnet} was used as the MD force calculator. Both the original (``vanilla'') CHGNet uMLIP and the version fine-tuned specifically for hcp Zn were applied. Fine-tuning was based on short relaxation trajectories \cite{Tolborg2023low} computed at the Density Functional Theory (DFT) level.

DFT calculations were performed with the Perdew--Burke--Ernzerhof (PBE) exchange-correlation functional \cite{PBE_1, PBE_2} as implemented in the VASP code \cite{vasp_code_1, vasp_code_2, vasp_code_3, vasp_code_4, vasp_code_PAW}. A kinetic energy cutoff of 520 eV and a $\Gamma$-centered $32  \times 32 \times 16$ $k$-point grid were used for the unit cell, with proportionally smaller grids for supercells. Methfessel--Paxton smearing \cite{MP_smearing}
with a 0.2~eV broadening was applied. In total, 35 relaxation trajectories with $2 \times 2 \times 1$ supercells (8 atoms) were calculated, covering 7 isotropic strain states from $-3\%$ to $+3\%$ in $1\%$ increments, and 5 trajectories with $4 \times 4 \times 2$ supercells (64 atoms) at equilibrium DFT volume. Each trajectory was initilized by randomly displacing atoms by 0.2~\AA\ and included 10 relaxation steps, yielding 400 DFT frames in total for CHGNet fine-tuning based on computed energies, forces, and stresses. Fine-tuning used the Adam optimizer as implemented in the CHGNet package \cite{deng2023chgnet}, running for five epochs with a batch size of 4 and an initial learning rate of $10^{-2}$. The training, validation, and test sets ratio was set to 8:1:1. This training achieved mean absolute errors (MAEs) of 1~meV/atom, 22~meV/\AA, and 0.07~GPa on the test set. Additional testing with $4 \times 4 \times 2$ supercells, where atoms were randomly displaced (using Gaussian-distributed displacements with a 0.1~\AA\ standard deviation), yielded similar MAEs within 0.8~meV/atom, 50~meV/\AA, and 0.10~GPa.

For the MD simulations, we employed an $8a \times 8a \times 4c$ supercell containing 512 atoms.
The cell parameters were fixed either to the experimental values
($a = 2.6636$~\AA, $c = 4.9457$~\AA) \cite{Nuss2010} or to those of the fully relaxed DFT structure ($a = 2.6445$~\AA, $c = 5.0461$~\AA). The simulation consisted of a 20~ps equilibration run followed by a 40~ps production run.  
A total of 8000 MD snapshots from the production run were used to compute the CA-EXAFS spectrum using the MD--EXAFS approach \cite{Kuzmin2009,KUZMIN2020rev}.   The EXAFS spectrum for each snapshot was calculated using the FEFF8.5L code \cite{Rehr2000,Ankudinov1998}, as described in Section\ \ref{RMC}.

\begin{figure}
	\centering
	\includegraphics[width=.9\columnwidth]{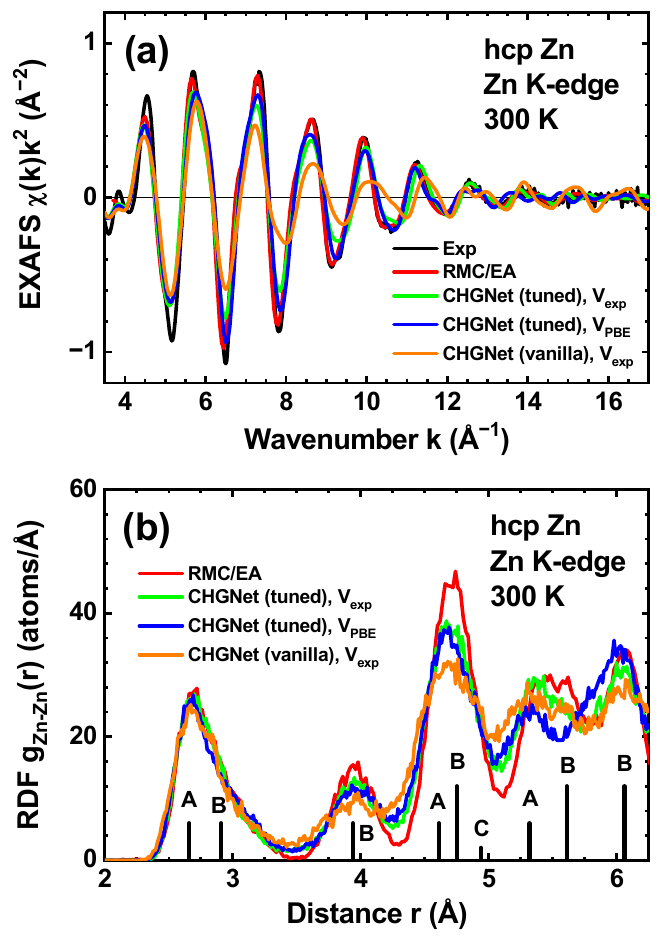}
	\caption{ (a) Experimental and calculated Zn K-edge EXAFS spectra $\chi(k)k^2$ for hcp zinc at 300~K. 
		(b) Radial distribution functions (RDFs) $g_\text{Zn--Zn}(r)$\ in hcp zinc at 300~K obtained from RMC/EA simulations based on the experimental EXAFS spectrum and from NVT-MD  simulations using the original (``vanilla'') and fine-tuned CHGNet potentials. Vertical bars indicate the positions of the coordination shells as determined by diffraction. The shells are labeled by atom type according to Table\ \protect\ref{table1}. See text for details. }
	\label{fig7}
\end{figure}

The obtained results are shown in Fig.\ \ref{fig7}. The EXAFS spectra calculated using the original (``vanilla'') and fine-tuned CHGNet uMLIPs are compared in Fig.\ \ref{fig7}(a) with the experimental and RMC/EA spectra.  
Note that simulations using the fine-tuned CHGNet uMLIP were performed for two slightly different unit cell volumes, experimental ($V_{\rm exp}$) and fully relaxed DFT ($V_{\rm PBE}$), whereas the simulation using the 
``vanilla'' CHGNet uMLIP was performed only for the experimental ($V_{\rm exp}$) unit cell volume. 
The corresponding Zn--Zn RDFs, calculated from the coordinates of atoms in the supercell, are compared in Fig.\ \ref{fig7}(b). 

\section{Results and discussion}

The anisotropy of the hcp zinc structure is clearly visible in Fig.\ \ref{fig1}, where a fragment with a stacking sequence of layers along the $c$-axis is shown. For the interpretation of the Zn K-edge EXAFS spectra, the atoms surrounding the absorbing atom are divided into three groups. Zinc atoms located in the same plane as the absorbing atom are classified as A-type; two zinc atoms located above and below the absorber along the $c$-axis are classified as C-type; and all other atoms are classified as B-type. The contributions of the different atom types to the first nine coordination shells are given in Table\ \ref{table1}.

The temperature dependences of the experimental EXAFS spectra, $\chi(k)k^2$, of hcp zinc and their Fourier transforms are shown in Fig.\ \ref{fig2}. The experimental data are of excellent quality, and peaks of structural origin are clearly observed in the FTs up to 10~\AA. An increase in temperature from 10 to 300~K results in strong damping of the EXAFS oscillations at high $k$ values (Fig.\ \ref{fig2}(a)) due to increased atomic thermal vibrations, which contribute to the mean-square relative displacement (MSRD). This effect is also observed in the Fourier transforms as a decrease in peak amplitudes (Fig.\ \ref{fig2}(b)). The contributions from the nearest nine coordination shells around the absorbing zinc atom (Table\ \ref{table1}) are  responsible for the peaks up to about 6~\AA\  in FTs.  

Deciphering of structural information from such a complex EXAFS spectrum is a challenging problem because of the large number of contributing scattering paths \cite{Kuzmin2014}. The analysis, restricted to a maximum half-path length of 6.5~\AA, indicates that there are 9 single-scattering paths, 41 double-scattering paths, and  28 triple-scattering paths in hcp zinc, i.e., 78 paths in total.  At the same time, based on the Nyquist criterion \cite{Stern1993}, a reasonable estimate of the total number of independent parameters for the data reported in Fig.\ \ref{fig2} is 
$N_{\rm max} = (2 \Delta k \Delta R) / \pi = 69$. This requires neglecting many paths and thus leads to a significant decrease in accuracy.   A more robust approach to the analysis uses atomistic simulations.  Here, we employed the RMC method \cite{Timoshenko2012rmc,Timoshenko2014rmc}, which has been shown to work well for both simple \cite{Jonane2018} and multicomponent  \cite{Bakradze2023,Drescher2025} compounds. 

The results of the RMC simulations for four representative temperatures (10, 100, 200, and 300~K) are shown in Fig.\ \ref{fig3} and indicate good agreement in both $k$- and $R$-space.  The atomic coordinates of zinc atoms in the final simulation box were used to calculate the RDFs $g_\text{Zn--Zn}(r)$ (Fig.\ \ref{fig4}).  Note that the peak positions in the RDFs agree with the coordination shell positions determined by diffraction. One can also see that the peaks corresponding to shells composed of zinc atoms located in the same plane as the absorber (A-type) vary less with increasing temperature  due to stronger interatomic  interactions (Table\ \protect\ref{table1}).

The RDFs were decomposed into a set of Gaussian functions corresponding to the first nine coordination shells (see Supplementary Material), and the MSRD factors $\sigma_{ij}^2$ for eight shells, composed of A and B type zinc atoms, were obtained as the variances of the distribution.
Note that the MSRD $\sigma_{ij}^2$\ is related to the mean-square displacements (MSDs) of atoms $i$ and $j$, usually probed in diffraction experiments, as $\sigma_{ij}^2 = MSD_i + MSD_j - 2 \phi \sqrt{MSD_i} \sqrt{MSD_j}$, where $\phi$ is a dimensionless correlation parameter \cite{Booth1995,Jeong2003}.

Temperature dependences  of the MSRDs $\sigma^2(T)$\  for the eight coordination shells of zinc are reported in Fig.\ \ref{fig5}(a). Note that the MSRDs of the 6th coordination shell (C-type) are missing because the contribution
from only two atoms, located at 4.94~\AA\ (Table\ \ref{table1}), is weak and cannot be reliably determined. 
These temperature dependences were approximated by the correlated Einstein model \cite{Sevillano1979}
\begin{equation}
	\label {Einstein}
	\sigma^2(T)={\frac{\hbar}{2\mu\omega_E}\coth\frac{\hbar\omega_E}{2k_BT}},
\end{equation}
where $\mu$ is the reduced mass of the atomic pair, $\omega_E$ is the Einstein frequency, and $k_B$ is the Boltzmann constant. The effective force constants $\kappa=\mu\omega_E^2 $ were calculated and are reported in Table\ \ref{table1}. A-type atoms,  located in the same plane, have the largest values of $\kappa$, indicating stronger interactions between them, which are slightly reduced at long distances due to the loss of correlation in atomic motion.  B-type atoms, located in layers different from that of the absorbing atom, have significantly smaller (nearly twofold) $\kappa$\ values, reflecting the anisotropy of the structure and lattice dynamics. Since the MSRD values for C-type atoms are not available, we instead use the values for a closely located group of B-type atoms in the second coordination shell  ($r_2({\rm {Zn-Zn}})=2.91$~\AA). The MSRD values of these atoms are among the largest, being only slightly smaller than those of the 7th coordination shell.

In Fig.\ \ref{fig5}(b), we compile information on the temperature dependence of the MSDs obtained from different experiments \cite{Skelton1968,Rossmanith1978,Nuss2010,Masadeh2022} and calculated in \cite{Peng1996} from the experimentally determined phonon density of states. 
The anisotropy of thermal vibrations is clearly evidenced by the difference between the out-off-plane MSD ($U_{33}$) and in-plane MSD ($U_{11}$) can be quantified using their ratio $U_{33}/U_{11}$ \cite{Nuss2010}.

The MSRD dependence on the interatomic distance is shown in Fig.\ \ref{fig6}(a). At large distances, when the correlation of atomic motion is reduced, it is expected to reach a constant value equal to 2MSD \cite{Jeong2003}. 
Deviations from this value depend on the importance of correlation effects, which are determined by interatomic interactions and the crystallographic direction, i.e., structural anisotropy.  Such effects are observed in the two 
nearest shells, located at  2.66~\AA\ and 2.91~\AA. Strong correlation of atomic motion in the first coordination shell is due to direct chemical bonding, whereas the reduced correlation in the second shell is caused by structure anisotropy, leading to much weaker interactions between atoms located in adjacent layers (Fig.\ \ref{fig1}).   

The temperature dependence of the anisotropy, expressed by the ratio of the MSRD factors MSRD$_2$/MSRD$_1$ in the second (MSRD$_2$) and first (MSRD$_1$) coordination shells,  is compared in Fig.\ \ref{fig6}(b) with the ratio of the MSD parameters  $U_{33}/U_{11}$ from \cite{Nuss2010}. Both dependences exhibit similar behavior, indicating an increase in anisotropy with increasing temperature by approximately 1.5--2 times, confirming the findings in \cite{Nuss2010}. These ratios indicate that the displacements of Zn atoms along the $c$-axis are approximately 1.2-1.6 times larger than those along the $a$-axis, consistent with the relatively large $c/a$ = 1.86 ratio for zinc compared with the ideal $c/a$ value.

The structural and bonding anisotropy of hcp zinc makes its atomistic simulations notoriously more challenging than those of simpler crystal structures with cubic lattices. Recent advances in the development of uMLIPs \cite{deng2023chgnet,batatia2023foundation} allow the computation of energies and forces with near ab initio accuracy, but at a significantly reduced computational cost. In particular, the CHGNet uMLIP \cite{deng2023chgnet} was recently used by us for MD-EXAFS simulations of the isostructural layered materials 2H$_c$-WS$_2$ and 2H$_c$-MoS$_2$, demonstrating good agreement with experimental EXAFS spectra after fine-tuning on a limited number of additional structures \cite{Zguns2025}. Here, we use a similar approach, described in Section~\ref{MD}. 

The original (``vanilla'') and fine-tuned CHGNet uMLIPs were used in NVT MD simulations performed at 300~K for two slightly 
different simulation box sizes: the experimental volume  ($V_{\rm exp}$) and the fully relaxed DFT volume ($V_{\rm PBE}$).
The experimental Zn K-edge EXAFS spectra are compared with the calculated ones in Fig.\ \ref{fig7}(a). 
The two EXAFS spectra, calculated  with the fine-tuned CHGNet uMLIP, show a weak but visible effect of 
box size on the spectral shape. This effect manifests mainly as a  difference in  the EXAFS phase, arising from variations in interatomic bond lengths. At the same time, the EXAFS amplitudes are close, indicating comparable levels of thermal disorder. 
Both calculated EXAFS spectra are in good agreement with the experimental and RMC-derived results. In contrast,
the EXAFS spectrum obtained using the original (``vanilla'')  CHGNet uMLIP deviates significantly from the other data in both amplitude and phase due to force underestimation (softening), which is a typical issue for uMLIPs \cite{Deng2025}. 
Thus, fine-tuning the potential model largely mitigates the softening effect and improves  agreement with the experiment  \cite{Zguns2025}. 

These conclusions are further supported by a comparison of the Zn--Zn RDFs $g_\text{Zn--Zn}(r)$, shown in Fig.\ \ref{fig7}(b). 
Since the RMC method provides a good fit to the experimental EXAFS data, the RDF obtained using RMC is considered a “reference” for MD simulations based on uMLIPs. While all simulations produce a similar shape for the first broad peak, located between 2.2 and 3.5~\AA\ and corresponding to the first two coordination shells (see also Fig.\ \ref{fig4}),  some differences appear for the outer coordination shells. The peaks in the RDF obtained using the original (``vanilla'')  CHGNet uMLIP are the most broadened as a result of the potential softening \cite{Zguns2025,Deng2025}.  At the same time, the RDFs calculated from the MD simulations using the fine-tuned CHGNet uMLIP are similar to each other and closer to the RDF obtained by the RMC method. 
However, the observed difference between radial distribution functions at large distances leaves room for further improvements in uMLIP.

\section{Conclusions}

Temperature-dependent (10-300~K) Zn K-edge X-ray absorption spectroscopy was used to study the  anisotropy of lattice vibrations in hexagonal close-packed zinc. The analysis of the experimental EXAFS spectra  
using reverse Monte Carlo simulations coupled with an evolutionary algorithm approach allowed 
us to determine accurate information on the mean-square relative displacements (MSRDs) for Zn–-Zn atom pairs in the first eight coordination shells. 

The temperature dependences of the MSRDs were approximated using the Einstein model to compute the effective force constants (Table\ \ref{table1}). 
The anisotropic character of thermal vibrations was expressed by the ratio of MSRD$_2$/MSRD$_1$ in the second (MSRD$_2$) and first (MSRD$_1$) coordination shells of zinc. This result is in good agreement with the ratio of the MSD parameters $U_{33}$/$U_{11}$ determined using single-crystal X-ray diffraction in \cite{Nuss2010}. 

Finally, the lattice dynamics of hcp Zn was simulated using the original (``vanilla'') 
and fine-tuned CHGNet universal machine-learning interatomic potentials (uMLIPs). 
The EXAFS spectra were computed based on the results of molecular dynamics simulations performed 
with the two uMLIPS at 300~K. Good agreement between the experimental and calculated EXAFS spectra was 
found for the fine-tuned CHGNet uMLIP, whereas the simulations based on the ``vanilla'' potential resulted in a damped EXAFS spectrum  due to the broadening of the Zn--Zn RDF caused by uMLIP softening \cite{Zguns2025,Deng2025}. The RDF computed from the MD simulations with the fine-tuned CHGNet uMLIP is  also closer to the RDF obtained by RMC. Thus, fine-tuning of the CHGNet uMLIP significantly improves the agreement with the experiment. 
These findings highlight the importance of accurately capturing anisotropic interactions for modeling lattice dynamics in metals and establish a combined EXAFS–atomistic simulation framework as a robust approach for validating and refining machine-learning interatomic potentials.

\printcredits

\section*{Declaration of Competing Interest}

The authors declare that they have no known competing financial interests or personal relationships that could have appeared to influence the work reported in this paper.

\section*{Acknowledgements}

This study was supported by the Latvian Council of Science project No. LZP-2022/1-0608.
P.\v{Z}. acknowledges the support of the project No. 1.1.1.9/LZP/1/24/016 by European Regional Development Fund.
We acknowledge DESY (Hamburg, Germany), a member of the Helmholtz Association HGF, for the provision of experimental facilities. 
Parts of this research were carried out at PETRA III and we would like to thank Dr. Edmund Welter for his assistance in using the P65 beamline. 
Beamtime was allocated for the proposal I-20211234 EC.

\section*{Appendix A. Supplementary material}\label{Supplementary}

Supplementary material

\section*{Data availability statement}

Data will be made available on request.



\end{document}